\documentclass[11pt]{article}

\usepackage[version=3]{mhchem} 
\usepackage{amsmath}
\usepackage[colorlinks = true,
linkcolor = blue,
urlcolor = blue,
citecolor = blue,
anchorcolor = blue]{hyperref}
\usepackage{pifont}
\usepackage{ulem}
\usepackage{multirow}
\usepackage{caption}
\usepackage[table,xcdraw]{xcolor}
\usepackage{hhline}
\usepackage{bbold}
\usepackage{setspace} 
\usepackage[document]{ragged2e}

\usepackage{etoolbox}
\usepackage{authblk}

\usepackage{graphicx} 
\usepackage{amsmath} 
\usepackage{amssymb} 
\usepackage{hyperref} 
\usepackage{authblk} 
\usepackage{float}
\usepackage{geometry}
\usepackage{jabbrv}

\DefineJournalAbbreviation{Reports}{Rep} 
\DefineJournalAbbreviation{Computational}{Comp}
\DefineJournalAbbreviation{Materials}{Mat}
\DefineJournalAbbreviation{Advances}{Adv}
\DefineJournalAbbreviation{Letters}{Lett}
\DefineJournalAbbreviation{Technology}{Tech}

\usepackage[style=phys]{biblatex}
\appto{\bibsetup}{\raggedright}

\usepackage{xr}
\newcommand{\Ang}{\text{\AA}}
\newcommand{\dwell}{$\mu$s per pixel}
\newcommand{\degC}{$^\circ$C}

\newcommand{\mc}{\multicolumn}
\newcommand{\cc}{\cellcolor[HTML]{DADADA}}
\newcommand{\tbf}[1]{\textbf{#1}}

\title{Leveraging Generative Adversarial Networks to Create Realistic Scanning Transmission Electron Microscopy Images}

\author[1,+]{Abid Khan}
\author[2,+]{Chia-Hao Lee}
\author[2,3,]{Pinshane Y. Huang \thanks{\href{mailto:pyhuang@illinois.edu}{pyhuang@illinois.edu}}}
\author[1,4,]{Bryan K. Clark \thanks{\href{mailto:bkclark@illinois.edu}{bkclark@illinois.edu}}}

\affil[1]{Department of Physics, University of Illinois Urbana-Champaign, Urbana, Illinois, 61801, United States}
\affil[2]{Department of Materials Science and Engineering, University of Illinois Urbana-Champaign, Urbana, Illinois, 61801, United States}
\affil[3]{Materials Research Laboratory, University of Illinois Urbana-Champaign, Urbana, Illinois, 61801, United States}
\affil[4]{Institute for Condensed Matter Theory and IQUIST and NCSA Center for Artificial Intelligence Innovation, University of Illinois Urbana-Champaign, Urbana, Illinois, 61801, United States}
\affil[+]{These authors contributed equally to this work}

\date{}
\begin{document}

\maketitle
\newpage

\begin{abstract} 
The rise of automation and machine learning (ML) in electron microscopy has the potential to revolutionize materials research through autonomous data collection and processing. A significant challenge lies in developing ML models that rapidly generalize to large data sets under varying experimental conditions. We address this by employing a cycle generative adversarial network (CycleGAN) with a reciprocal space discriminator, which augments simulated data with realistic spatial frequency information. This allows the CycleGAN to generate images nearly indistinguishable from real data and provide labels for ML applications. We showcase our approach by training a fully convolutional network (FCN) to identify single atom defects in a 4.5 million atom data set, collected using automated acquisition in an aberration-corrected scanning transmission electron microscope (STEM). Our method produces adaptable FCNs that can adjust to dynamically changing experimental variables with minimal intervention, marking a crucial step towards fully autonomous harnessing of microscopy big data.
\end{abstract}

\section*{Keywords}
Cycle Generative Adversarial Network (CycleGAN), deep learning, scanning transmission electron microscopy (STEM), 2D materials, single-atom defects




\section*{INTRODUCTION} 

Machine learning (ML) techniques have been widely applied in electron microscopy for applications such as atom localization~\cite{Ziatdinov2017a, Madsen2018, Lin2021a}, defect identification~\cite{Maksov2019, Lee2020, Guo2021}, image denoising~\cite{Quan2019, Ede2019a, Wang2020e}, determining crystal tilts and thickness~\cite{Xu2018, Zhang2020, Yuan2021}, classifying crystal structures~\cite{Aguiar2019, Kaufmann2020}, optimizing convergence angles~\cite{Schnitzer2020}, identifying Bragg disks~\cite{Munshi2022}, visualizing material deformations~\cite{Shi2022}, automated microscope alignment~\cite{Xu2022}, and many others. Several recent reviews~\cite{Ede2020, Kalinin2022, Botifoll2022} provide an overview of new and emerging opportunities at the interface of electron microscopy and ML. 

One of the biggest challenges of ML in materials research is that it requires a large amount of high quality training data, which must be paired with the ground truth for supervised learning methods. For example, training a defect identification network requires both a set of images and a known set of defect positions in each image. Because manual labelling is extremely time consuming and prone to human bias and error, the typical approach is to train ML models using simulated data, which simultaneously produces both images and labels. However, simulated scanning transmission electron microscopy (STEM) images deviate non-trivially from experimental images because it is extremely difficult to accurately reproduce complex experimental factors that impact the image, including detector noise~\cite{Seki2018, Jones2016}, sample drift and scan distortions~\cite{Braidy2012, Ophus2016, Savitzky2018}, time-dependent alignment errors~\cite{Schramm2012}, radiation damage~\cite{Egerton2013}, and surface contamination~\cite{Hettler2017, Goh2020}. These factors make it challenging to train high-accuracy ML models with simulated data; often times such training requires considerable manual optimization to generate a usable training set for a specific experiment data set. Most of these experimental imperfections can be simulated to certain degree \cite{Lee2020, Lin2021a}, but doing so accurately requires significant parameter optimization for each set of experimental conditions. To maximize their accuracy and precision, ML models are typically trained using a small distribution of simulation parameters, which are carefully chosen to cover most of the experimental variations. Such models must be re-trained each time the effective resolution or contrast of the image changes, which may occur several times a day, even during a single experimental session. Far from realizing the promise of fully automated image processing, this lack of generalizability of ML models represents a significant barrier for microscopy data processing at the scale of big data.

Here, we construct a cycle generative adversarial network (CycleGAN)~\cite{Zhu2017} to minimize the difference between simulated and experimental STEM data, producing realistic training data while simultaneously preserving the ground truth for ML applications. CycleGANs are often used for image-to-image translation, such as style transfer tasks like converting a photograph into a Monet-like painting, or changing an image of a zebra into a horse~\cite{Zhu2017}. We utilize CycleGANs to generate a high quality, CycleGAN-processed training set for defect identification tasks using a fully convolutional network (FCN)~\cite{Ziatdinov2017a, Madsen2018, Maksov2019, Lee2020}. We find that FCNs trained on CycleGAN-processed images show comparable defect identification performance with the manually optimized training set, while requiring much less human intervention.


\section*{RESULTS AND DISCUSSIONS}
\subsection*{Generating realistic images with CycleGANs}
\begin{figure}[H]
    \centering
    \includegraphics[width = 16 cm]{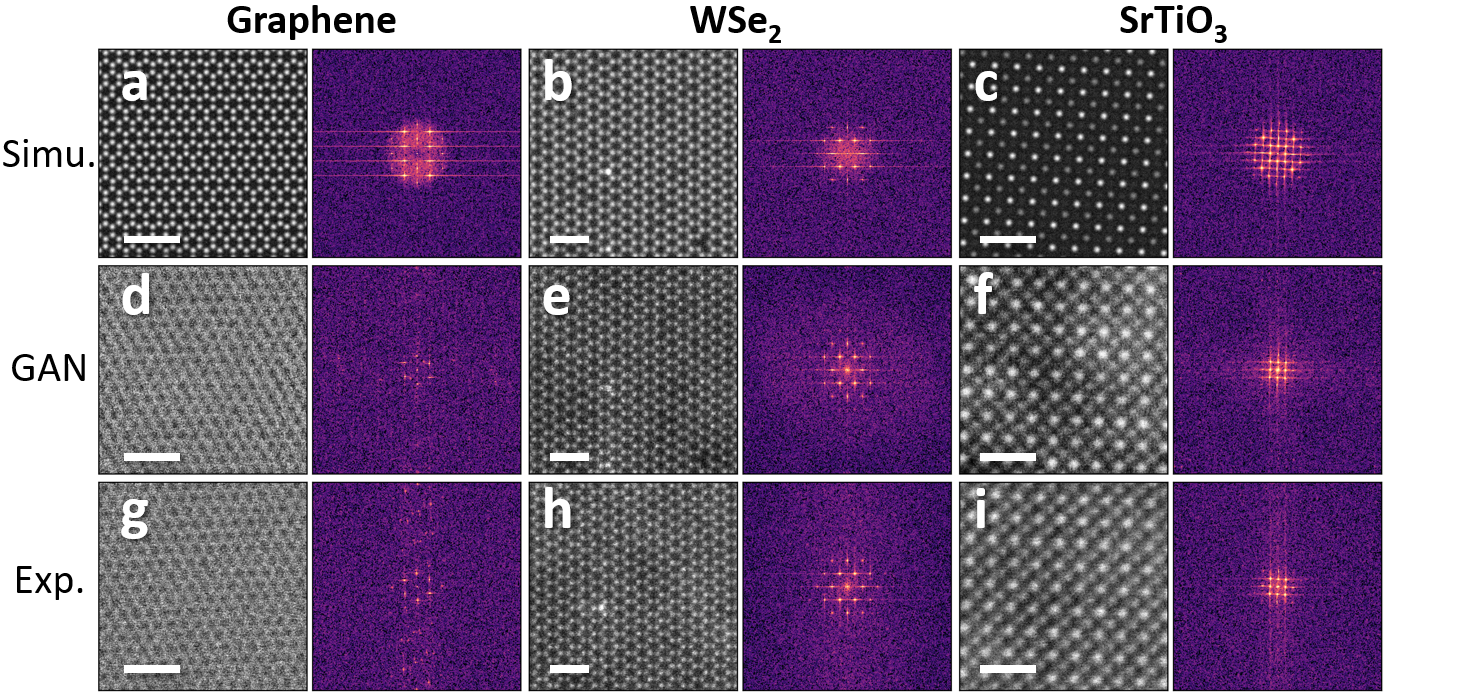}
    \caption{Representative images and their power spectra of simulated, CycleGAN-processed, and experimental images of graphene, \ce{WSe2}, and \ce{SrTiO3}. Each row displays the real space image along with its power spectrum. Power spectrum of the image is calculated by the fast Fourier transform (FFT) and displayed in log scale for clarity. (a--c) Semi-quantitative, simulated ADF-STEM images are generated by the Computem program~\cite{Kirkland2013a, Kirkland2020}. These simulated images are passed into pre-trained CycleGANs and resulted in highly realistic images in (d--f). Individual CycleGANs are trained for each materials system using both simulated and experimental data as the training set. The CycleGANs transfer experimental imperfections, including noise, jittering, distortion, and surface contamination to simulated images, making the simulated images appear realistic. (g--i) Experimental ADF-STEM images and their power spectra for comparison.  Note that (h) has been selected to contain the same type of bright defect as in (e) so that images are more comparable. Scale bars are equal to 1 nm.}
    \label{fig:1}
\end{figure}

Figure~\ref{fig:1} shows an array of simulated (Figure~\ref{fig:1}a--c), CycleGAN-processed (Figure~\ref{fig:1}d--f), and experimental (Figure~\ref{fig:1}g--i) annular dark field (ADF) STEM images and their power spectra for three different materials: monolayer graphene, monolayer \ce{WSe2}, and bulk \ce{SrTiO3} (see Methods for experimental details). The simulated images are calculated with the `incoSTEM' package in Computem~\cite{Kirkland2013a, Kirkland2020} with minimal parameter optimization. These out-of-the-box simulated images are a poor qualitative and quantitative match for the experimental data for two main reasons. First, in contrast with quantitatively accurate multislice or Bloch wave approaches for STEM image simulation, incoherent approximations such as incoSTEM are mainly used because they output images nearly instantaneously, and are thus mainly used to generate initial, qualitative image simulations. Second, images simulated using any method lack key aspects of real experimental data, which contain detector noise, drift-induced distortions, probe jittering, lens aberrations, thickness variations, and surface contamination or damage. 

Next, we generate realistic STEM images (Figure~\ref{fig:1}d--f) by feeding the simulated images (Figure~\ref{fig:1}a--c) into the CycleGAN. We trained a separate CycleGAN for each material system, because CycleGANs are typically unable to perform image translation when large shape changes are needed~\cite{Gokaslan2018}. The CycleGAN-processed images are similar to the experimental ones in both real and reciprocal space, including their atom size, noise level, image contrast, and contamination. These CycleGAN-processed images can be used as high-quality ML training data. The use of CycleGAN removes the need for manual parameter optimization in the simulation process, but unlike conventional GAN, preserves the original labels and associated ground truth. The CycleGANs are trained with both experimental and out-of-the-box simulated images (see Methods for training details) and do not require paired training data. This makes CycleGAN particularly useful for this `simulation-to-experiment' style transfer task, because a large number of simulated images can be generated without manually matching all atom positions with the experimental data. 

\subsection*{CycleGAN architecture and optimization}
\begin{figure}[H]
    \centering
    \includegraphics[width = 16 cm]{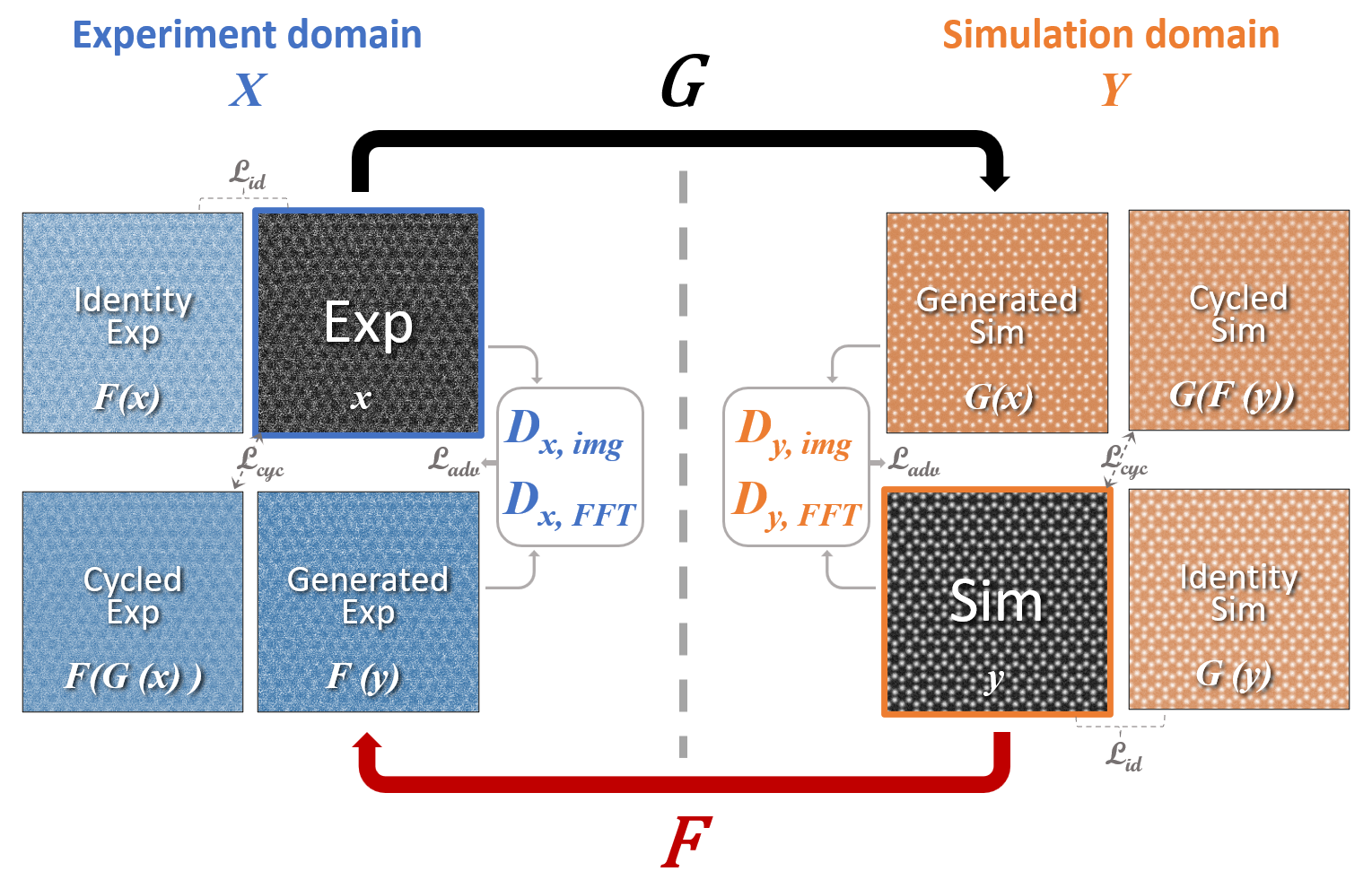}
    \caption{Schematic of the major components in a CycleGAN. The CycleGAN translates images between two domains: (1) the experiment domain $X$ and (2) the simulation domain $Y$, and has two main components: generators and discriminators. The generator $G$ converts input experimental images $x$ to simulation-like images $G(x)$, and the generator $F$ converts input simulated images $y$ to experiment-like images $F(y)$. The generated images ($F(y)$ and $G(x)$) are then fed into four discriminators ($D_{x, \text{img}}$, $D_{x, \text{FFT}}$, $D_{y, \text{img}}$, $D_{y, \text{FFT}}$) along with the raw images ($x$ and $y$) to evaluate the quality of generated images. Both the input images and their FFTs are examined by the discriminators to calculate the adversarial losses ($L_{\text{adv}}$), which are used to optimize both generators $F$ and $G$ respectively. By passing the raw images ($x$ and $y$) with combinations of both generators, identity images ($F(x)$ and $G(y)$) and cycled images ($F(G(x))$ and $G(F(y))$) are also generated. The corresponding identity loss $L_{\text{id}}$ and cycle consistency loss $L_{\text{cyc}}$ are added to ensure the identity and cycle consistency mapping of the generators.}
    \label{fig:2}
\end{figure}

Figure~\ref{fig:2} shows the `cycle' structure of our CycleGAN. In our work, the CycleGAN converts the images between experiment domain $X$ and simulation domain $Y$. The conversion between domains is commonly called a `mapping.' The CycleGAN provides two mappings, $G: X \rightarrow Y$, which makes experimental images look like simulations; and $F: Y \rightarrow X$, which makes simulated images look experimental.

To implement this pair of mappings, the CycleGAN is built on two separate GANs connected in a cyclic manner, where each GAN is responsible for converting images from one domain to the other domain. The main components of our CycleGAN are neural networks called the `generator' and the `discriminator'. The goal of a generator is to generate high-quality `fake' images from inputs of another domain, while the goal of a discriminator is to distinguish whether the output from the generator is real or fake. Hence, the training of a CycleGAN is essentially the competition between generators and discriminators. At the end of the training, the final products are two well-trained generators that act as the domain-to-domain mappings $G$ and $F$. Here, we mainly use the CycleGAN to map simulated images into experiment-like images. However, as we show in Supplemental Figure1, the generator $G$ can also be used to effectively remove imperfections from experimental data to produce simulation-like, high signal-to-noise ratio images that can be more easily interpreted through other data-processing methods.

Our CycleGAN contains two generators ($G$ and $F$), two discriminators in experiment domain $X$ ($D_{x, \text{img}}$ and $D_{x, \text{FFT}}$), and two discriminators in simulation domain $Y$ ($D_{y, \text{img}}$ and $D_{y, \text{FFT}}$) (Methods). While GANs typically operate in real space, we add Fourier space discriminators ($D_{x, \text{FFT}}$ and $D_{y, \text{FFT}}$) because certain information, such as low-frequency contamination and high-frequency noise, are better encoded in Fourier space. The generators are then forced to generate images that can fool not only the real-space image discriminator, but also the Fourier space discriminator. The FFT discriminators are critical for ensuring both the real- and k-space information of the generated STEM images look realistic, especially for images with lower signal-to-noise ratio. For example, a CycleGAN trained without the FFT discriminator produces streak artifacts in graphene images and their FFTs, as shown in Supplemental Figure2.

The optimization of the 6 neural networks ($G$, $F$, and $D$s) is done by minimizing a set of loss functions (Methods). CycleGANs contain additional machinery in the form of regularization terms to ensure the stability and reversibility of the two mappings $F$ and $G$. The first is the cycle-consistency loss $L_{\text{cyc}}(F,G)$, which regularizes the mappings so that a full cycle of forward and backward operation returns the input image back to its original domain and be similar to the initial input. Mathematically, $L_{\text{cyc}}$ enforces $F(G(x)) \approx x$, and $G(F(y)) \approx y$ to preserve major image information while transferring between domains. Additionally, the identity loss $L_{\text{id}}$ aims to have $F(x) \approx x$ and $G(y) \approx y$, so that images would be mapped to themselves if they are already in the target domain. Without these two constraints, it is likely that a trained mapping will map a simulated image into a realistic experiment-like image that will otherwise have no relation to the original input, i.e. it will not preserve important features like atomic defects.

\subsection*{Evaluation metrics of CycleGAN images}

\begin{figure}[H]
    \centering
    \includegraphics[width = 16 cm]{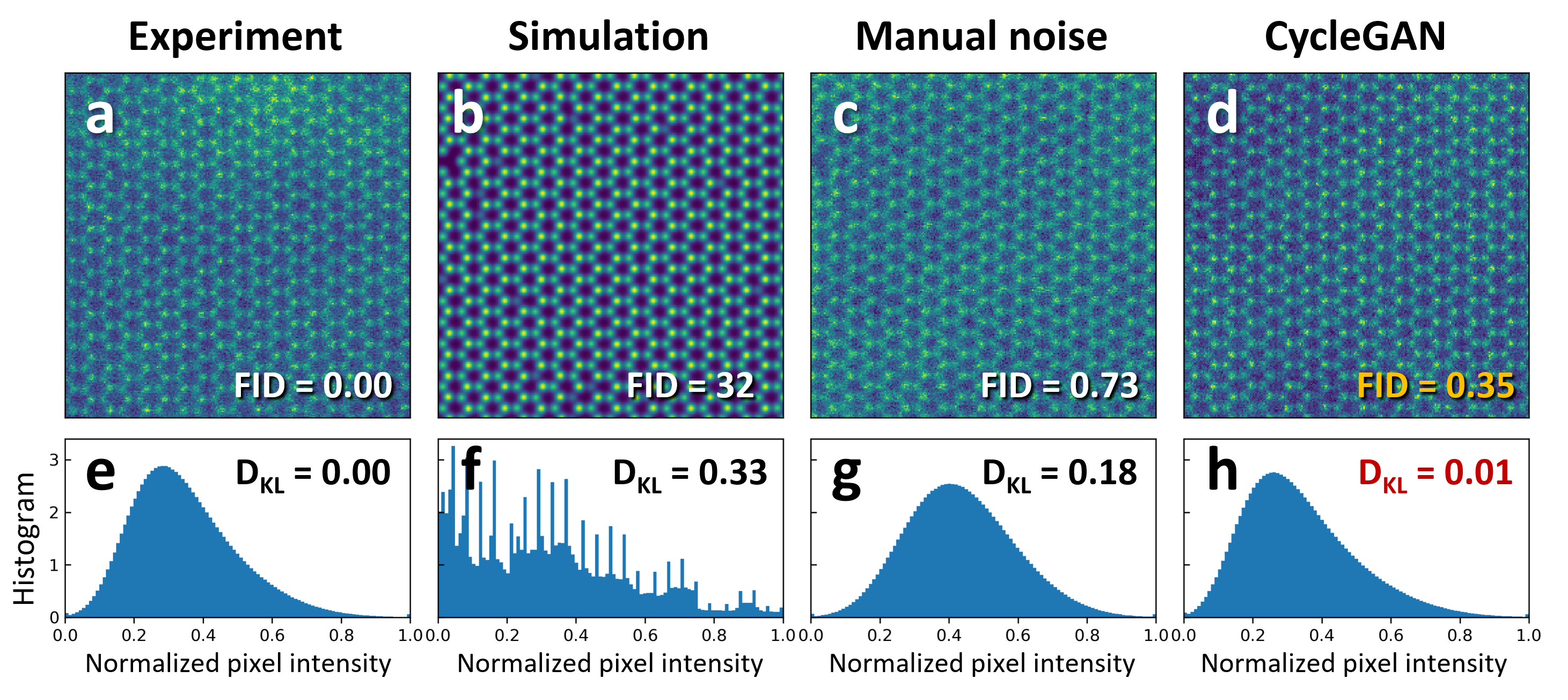}
    \caption{Quantitative measurements of data set quality using the Fr\'echet inception distance (FID) and the Kullback–Leibler (KL) divergence of \ce{WSe2}. Images are generated from (a) experiments, (b) simulation without noise, (c) simulation with manually optimized noise, and (d) CycleGAN. The FID score measures the dissimilarity between image data sets, so a smaller FID score implies higher similarity between image data sets. The FID scores with respect to the experimental data set are labeled at the bottom right corner of each representative image, with the lowest non-zero value marked in yellow. (e--h) Histograms of normalized pixel intensity are calculated for each image data set. Each histogram is normalized so that the probability distribution sums to unity. The KL divergence $D_{\text{KL}}(P||Q)$ of each data set with the experimental histogram are labeled as $D_{\text{KL}}$ at the top right corner of each histogram, with the lowest non-zero value marked in red. Note that both the FID score and KL divergence of intensity histograms are calculated for the entire data set with respect to the experimental data set, where each data set contains roughly 1,700 image patches with $256 \times 256$ pixels. The CycleGAN generated image set exhibits the best FID score and lowest KL divergence, indicating it is the best match for experimental data. }
    \label{fig:3}
\end{figure}

While Figure~\ref{fig:1} shows that CycleGAN-processed images are a good qualitative match with the experimental images, we also examined whether these image types are quantitatively indistinguishable by calculating the Fr\'echet Inception Distance (FID)~\cite{Heusel2017} and Kullback–Leibler (KL) divergences~\cite{Kullback1951} between CycleGAN and experimental images (Figure~\ref{fig:3}). The FID score is extensively used in the field of computer vision to evaluate the image quality and the performance of generative models~\cite{Lucic2018, Borji2019, Ren2020}. The FID score is evaluated by first converting the images from different data sets into feature vectors with a convolutional network (Inception v3) trained on the ImageNet~\cite{Heusel2017}, and then calculating the Fr\'echet distance~\cite{Dowson1982} between the feature vectors. Note that the feature vectors are extracted through a non-local and non-linear mapping using a convolutional network, specifically designed for computer vision tasks. This process captures high-level, emergent features of the STEM images and allows us to compare the similarity of two different image distributions. In general, a lower FID score indicates higher similarity and better performance of the generative model.

We calculate the FID scores of images generated from simulation without noise (Figure~\ref{fig:3}b), simulation with manually optimized noise (Figure~\ref{fig:3}c), and CycleGAN (Figure~\ref{fig:3}d). We choose experimental STEM images of \ce{WSe2} (Figure~\ref{fig:3}a) as our reference for the FID score calculation because the ultimate goal is to have a generator that outputs realistic, experiment-like images. We find that the CycleGAN-processed images show the lowest (i.e. best) FID score (FID = 0.35) compared to simulations without noise (FID = 32) and simulations with manually added noise (FID = 0.73). Table~\ref{tab:FIDs} shows the FID scores of graphene, \ce{WSe2}, and \ce{SrTiO3}. 

In addition to the FID score comparison, which involves an additional Inception model, we calculate the KL divergence directly from the pixel intensity histograms of different image data sets (Figure~\ref{fig:3}e--h). The KL divergence ($D_{\text{KL}}$) is a type of statistical distance that measures the difference between two probability distributions ($P$ and $Q$). It is computed as $D_{\text{KL}}(P||Q) = \sum P \log(\frac{P}{Q})$, so $D_{\text{KL}} \rightarrow 0$ when $P$ and $Q$ are similar. In other words, $D_{\text{KL}}(P||Q)$ is the amount of lost information when approximating $P$ with $Q$, therefore, it is widely used as a similarity metric for distributions. We find that the $D_{\text{KL}}$ of CycleGAN with respect the experimental histogram is significantly lower than the others, indicating the pixel intensity distribution is a better match to the experimental data than the other data sets. We also analyze the power spectra of the images and find that CycleGAN-processed images have the closest power distribution to experimental images (Supplemental Figure3), again indicating the high similarity between CycleGAN-processed and experimental images.

Finally, as a qualitative way to understand how the noise transfer works for CycleGANs, we train CycleGANs on one material system and apply it to another material system. This procedure allows us to separate the noise profile learned from the underlying crystal structure and visualize it (see Supplemental Figure 4).

\subsection*{Defect identification with CycleGAN and FCN}
We demonstrate the utility of the CycleGAN-processed, realistic STEM images for defect identification in a two dimensional (2D) material, monolayer \ce{WSe2}. Identifying atomic defects in atomic-resolution STEM data is critical to understanding the materials' structure-property relations. 2D materials are particularly well suited for this approach because they are atomically thin, making it possible to image individual atoms across large areas~\cite{Huang2013a, Lee2020}. Several studies~\cite{Krivanek2010d, Azizi2017, Zheng2019, Ding2021} have employed quantitative analysis of ADF-STEM image intensities, or so called Z-contrast, to identify atomic defects in 2D materials as the integrated intensity approximately scales with the atomic number $Z^{1.5-1.8}$. However, the integrated intensity is very sensitive to local contamination and microscope conditions, rendering this approach much less reliable for data sets with varying imaging conditions. 

\begin{figure}[H]
    \centering
    \includegraphics[width = 16 cm]{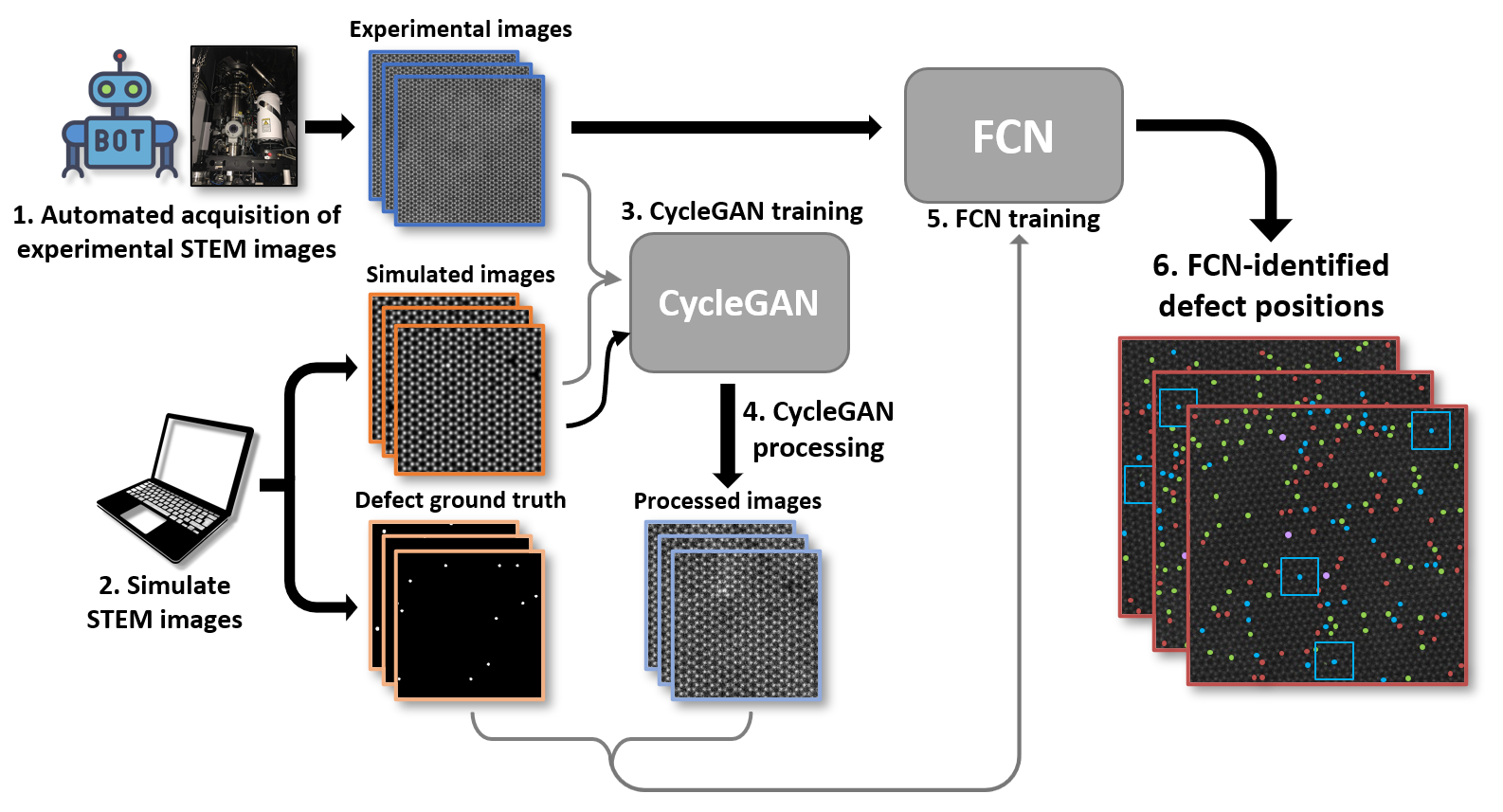}
    \caption{The workflow of detecting single-atom defects using both a CycleGAN and a fully convolutional network (FCN). First, we acquire large data sets of experimental STEM images with a custom-build automated acquisition script. We then simulate STEM images using Computem with pre-determined defect types and positions, which constitute the ground truth of defect labels. The experimental images and the simulated images are then used to train a CycleGAN, whose main purpose is to train a generator that can produce realistic images from simulated input images without altering the defect ground truth. CycleGAN learns the noise distribution from experimental images and adds it to the input simulated images, making them higher quality training data for further supervised learning applications. These CycleGAN-processed images, along with the defect ground truth, are used to train a defect identification FCN. Lastly, the FCN takes in the same experimental images with atomic defects, and returns the locations of those defects. The black arrows represent the input and output data flow, and the gray arrows indicate the input data for neural network training.}
    \label{fig:4}
\end{figure}

Figure~\ref{fig:4} shows our defect identification workflow. We utilize an FCN for defect identification; this FCN is trained on the CycleGAN-processed, realistic STEM images. The workflow has 6 main steps: (1) Acquire experimental images using automated scripts, (2) simulate STEM images, (3) train the CycleGAN with both experimental and simulated images, (4) use the CycleGAN to post-process simulated images into realistic images, (5) train the FCN with both CycleGAN-processed images from (4) and the corresponding defect ground truth from (2), and (6) identify atomic defects in experimental images with the FCN. Once set up, the workflow is relatively straightforward with steps being automatically linked together. These steps, and quantitative measurements of the resulting defect identification performance, are described in detail below.

\begin{figure}[H]
    \centering
    \includegraphics[width = 16 cm]{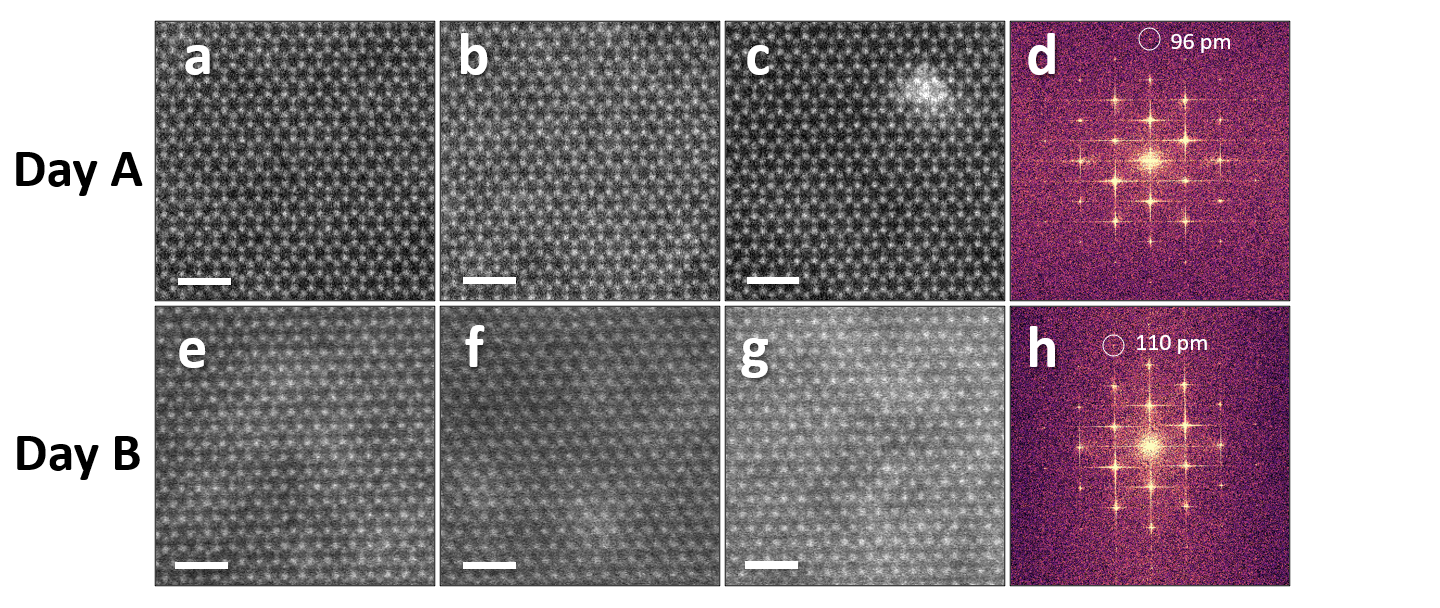}
    \caption{Example of variations within experimental STEM data sets taken on two different days, Day A and Day B. Experimental STEM images acquired on Day A (a--c) and B (e--g) showing variations within and between days due to sample contamination and microscope instabilities. (d and h) Power spectra obtained with FFT of (a and e) demonstrate the variation of image resolution, from 96 pm to 110 pm, determined by the highest transferred spatial frequency marked with white circles. The power spectra are displayed in log scale for visibility. Scale bars are equal to 1 nm.}
    \label{fig:5}
\end{figure}

First, large-scale experimental STEM images are automatically acquired with our custom-built script~\cite{Lee2022}. We acquire 2 experimental data sets of monolayer \ce{WSe2} on 2 different days. These data are labeled as `Day A' and `Day B', which include 107 and 211 images, respectively, with representative images shown in Figure~\ref{fig:5} (Methods). These data sets are not only massive (they contain 1.5 and 3 million atoms, respectively), but they also demonstrate the typical variations of imaging conditions on different days. The irregular, bright haziness on the atomic-resolution images corresponds to carbon-based surface contamination, whose spatial distribution, thickness, and intensity vary drastically between different regions in a single specimen. In addition, microscope instabilities can cause degradation of image resolution as shown in Figure~\ref{fig:5}d and h, where the resolution is degraded from 96 to 110 pm. In each of these data sets, we manually label single Se vacancy defects in 3 images to serve as the ground truth for FCN performance evaluation (Methods); this manual labeling is for testing purposes only and would not be needed in a typical workflow for data analysis. In total, 2 different test sets with 3 images each are constructed and labeled as `A' and `B'. By combining the experimental data sets (Day A and Day B) and test sets (A and B) separately, we obtain the 3\textsuperscript{rd} experimental data set (Day AB) and the 3\textsuperscript{rd} test set (AB). Note that `Day AB' contains $107 + 211 = 318$ images, while the test set `AB' has $3 + 3 = 6$ images.

\begin{figure}[H]
    \centering
    \includegraphics[width = 14 cm]{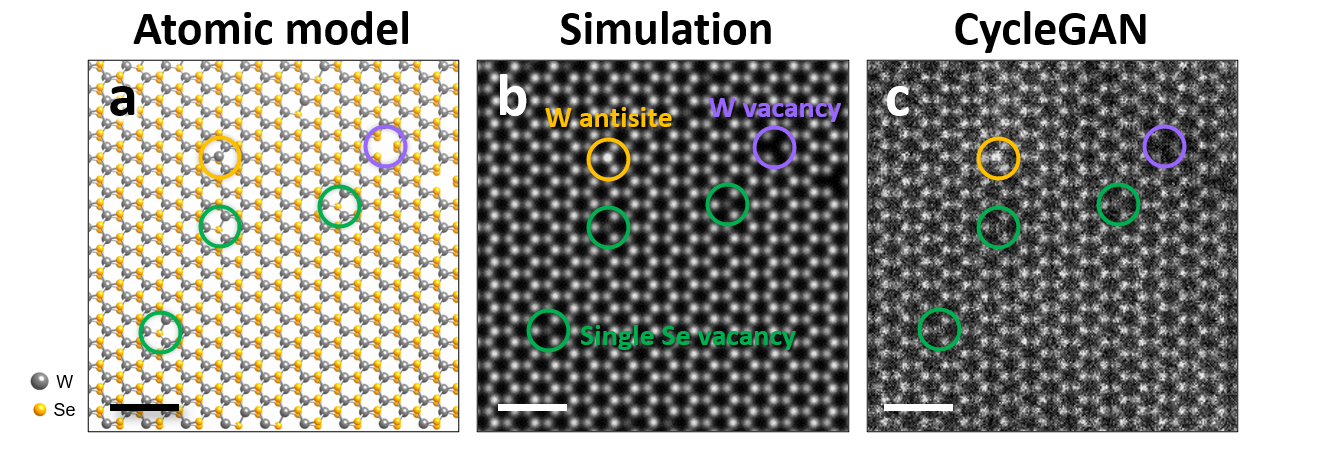}
    \caption{The CycleGAN-processed images preserve the defect types and positions from the input simulated images. (a) Atomic structure of \ce{WSe2} with atomic defects. The model is rotated by 10$^{\circ}$ for better visualization of the overlapping Se atoms and atomic defects. (b) Simulated images are generated by Computem with pre-determined defect types and positions, therefore the ground truth of defect labels are known. (c) By using the simulated images (b) as input, the CycleGAN-processed images show realistic, experimental-like noise distribution while keeping the defect label unaltered. Note that only 5 defects in the field of view are circled for better clarity. Scale bars are equal to 1 nm.}
    \label{fig:6}
\end{figure}

We then generate incoSTEM-simulated STEM images with defect labels, which contain the same atomic defect species with the experimental STEM images. Note that CycleGANs do not require that defect positions in the experimental and simulated images be identical for training, because it is an unsupervised learning technique for style-transfer between unpaired images~\cite{Zhu2017}. Here, the `style' being transferred is the `local texture', or noise, rather than the defects; the simulated images are generated with random defect distributions. Using both the experimental and simulated STEM images, we train the CycleGAN to generate realistic, high-quality simulated training data that matches the experimental conditions, while keeping the initial defect positions unchanged. Figure~\ref{fig:6} shows the defect positions before and after the CycleGAN processing. The defect positions (circled) are preserved by the CycleGAN because only the `local texure', or noise, is being transferred to the simulated images. This allows us to train an FCN with the combination of CycleGAN-processed STEM images and corresponding defect labels that are produced in the initial simulation. Once the CycleGAN is trained, it can generate unlimited amount of experiment-like training data because the sole input is the labeled out-of-the-box simulated data, which can be generated almost instantaneously.

\subsection*{FCNs performance with different training sets}
Lastly, we use the CycleGAN-generated images to train FCNs and evaluate their performance for defect identification on experimental images (Methods). Table~\ref{tab:FCNresults} shows the FCN performance with different FCN training sets. To demonstrate the flexibility of our CycleGAN approach, we trained 5 FCNs with 5 different training sets and evaluate their performance with 3 manually labeled experimental test sets (A, B, and AB) as described above.

The 5 different FCN training sets are: (1) simulation without noise or any processing `no noise', (2) simulation with manually optimized noise `manual noise', as described by our previous work~\cite{Lee2020}, (3) CycleGAN-A-processed images `CycleGAN-A', (4) CycleGAN-B-processed images `CycleGAN-B', and (5) CycleGAN-AB-processed images `CycleGAN-AB'. In this naming convention, CycleGAN-A-processed images means that the CycleGAN is trained with the experimental data set from day A. The FCN training sets are kept the same size, and other training details are described in the Methods. The FID scores of these training sets are described in Supplemental Table 1. Conventional classifier metrics including precision, recall, and F1 scores~\cite{Murphy2012} are chosen to evaluate the trained FCNs. Higher values indicate better FCN performance. While FCNs are typically tested with simulated data, FCNs that perform well on simulated data do not necessarily perform well on experimental data. For example, we show in Supplemental Table 2 that all of our FCNs exhibit high precision, recall, and F1 scores $(> 96 \%)$ when tested on simulated data. Here, we use a more challenging test for the FCNs: evaluating their performance directly on manually labeled experimental data.

When evaluated on experimental data, we observe excellent FCN performance for the CycleGAN processed data. As shown in Table~\ref{tab:FCNresults}, the FCNs trained with CycleGAN-processed images perform comparably well with the ones trained with manually optimized simulated images, while requiring significantly less human intervention. Note that precision and recall values above 90\% are considered exceptional, and often only achieved while training and testing with simulated data. 

We obtain the best results when we test the FCN on the same dataset of images that the CycleGAN is trained on (gray cells in Table~\ref{tab:FCNresults}). This suggests the CycleGAN is adapting to the microscopy conditions of that day. To further validate this, we train the CycleGAN on one day (i.e. data set) and then test the FCN using images from a different day. Here we should expect poor results due to the differing microscopy conditions and this is often what we find (the non-gray cells in Table~\ref{tab:FCNresults}) giving further evidence that the CycleGAN is learning local-in-time microscopy conditions. In comparison, the performance of the FCN trained with noise-free images (first row of Table~\ref{tab:FCNresults}) is significantly worse than the other 4 FCNs . This suggests that adding proper noise into the simulated images is necessary for the FCNs to correctly locate defects in experimental STEM images. This is not surprising because one would expect that FCNs perform the best when the input is well-described by the training set.
 
While we use 107 (Day A), 211 (Day B), and 318 (Day AB) experimental images for the CycleGAN training for the tests described above, we often achieve comparable FCN performance using as few as only 6 experimental images, though these small training sets occasionally lead to poor FCN performance (Supplemental Table 3). Notably, this small number of experimental images required for CycleGAN training opens the door to dynamically evolving ML models. Although training a CycleGAN from scratch takes 6 hours in our case, updating a pre-trained CycleGAN with a small set of new experimental images may take significantly less time, which could enable nearly in-time adaptation as the microscope conditions or sample properties evolve. While we chose FCN and defect identification as a simple example to demonstrate the application of CycleGAN, it is also possible to combine CycleGAN with more advanced architectures such as ensemble learning~\cite{Roccapriore2022} to achieve even higher generalization capacity for real-time microscopy applications.

In conclusion, we developed a CycleGAN approach to generate high quality, realistic STEM images and benchmarked its performance against manually optimized simulated images. Using this approach, we show that an FCN trained on CycleGAN-generated data achieves high precision for identifying atomic defects in STEM images, enabling flexible, high throughput data processing for large, automatically acquired experimental data sets. Our CycleGAN approach could potentially be applied to most imaging techniques with a proper forward model for generating realistic images for ML applications in microscopy, such as autonomous collection and processing of atomic resolution materials data~\cite{Spurgeon2020, Olszta2022, Kalinin2022}. 

\section*{METHODS} 
\subsection*{STEM sample preparation}
The monolayer graphene (Grolltex) was wet-transferred from a Cu foil to an \textit{in situ} heating chip (Protochips). The monolayer \ce{WSe2} was first mechanically exfoliated from bulk crystals (2D Semiconductors) using the Au-assisted exfoliation method~\cite{Velicky2018}, and then was transferred onto perforated TEM grids (Quantifoil) using a polymer free transfer~\cite{Pacil2008} modified with subsequent KOH and deionized water bath after the grid contact. The \ce{SrTiO3} sample was fabricated using the standard focused ion beam lift-out procedures in an FEI Helios 600i Dual Beam FIB-SEM system.

\subsection*{STEM image acquisition}
ADF-STEM image acquisition was conducted in a Thermo Fisher Scientific Themis-Z aberration-corrected STEM operated at 80 kV. For atomic-resolution ADF-STEM imaging, the point resolution was about 1 \Ang\ with 25 mrad convergence semi angle, 35 pA probe current, 63 to 200 mrad collection semi angles for \ce{WSe2} and \ce{SrTiO3}, 14--20 pm pixel size and a total dwell time of 20 \dwell\ using 10-frame averages. The STEM images for graphene were acquired with 100 pA, a 25 mrad inner collection angle, and were imaged after annealing at 1000 \degC\ for a few minutes to reduce the contamination.

\subsection*{STEM image simulation}
The STEM simulation is done with the `incoSTEM' package in Computem~\cite{Kirkland2013a, Kirkland2020} with minimal parameter optimization. The `Simulation', or `no noise' images in Figure~\ref{fig:3}b and Table~\ref{tab:FCNresults} are simulated with the same experimental conditions described above, except that no counting noise, source size, or aberrations are included. The manually optimized parameters for the `manual noise' data set in both Figure~\ref{fig:3}c and Table~\ref{tab:FCNresults} include surface contamination extracted from experimental images, Poisson and Gaussian noise, image shear, source size, aberrations up to 2\textsuperscript{nd} order, and image brightness/contrast. All atom positions are randomly perturbed (std = 0.01 \AA) before simulation to remove a periodic sampling artifact that can occur when atoms are not exactly located on a pixel, which would show up as stripes in their FFTs.

\subsection*{CycleGAN architecture}
The generators follow a U-net architecture as described in the appendix of Ref.~\cite{Isola2017}. The discriminators are patch-level~\cite{Zhu2017}, meaning that instead of the output being a single number indicating whether an image is real or fake, the discriminator outputs a $30 \times 30$ array of outputs, where each element in the array denotes the realness of a $70 \times 70$ patch of the original image. One difference between the architecture here and that in Ref.~\cite{Isola2017} is that we use an instance normalization instead of a batch normalization in each layer. The FFT discriminators are implemented using the log of the power spectrum of the image, or $\log(|FT(I)|^2)$, where $I$ is the image, and $FT$ is the Fourier transform. Therefore, only the amplitude information are being considered in the FFT discriminators. It is possible to extend the approach to complex values for phase information~\cite{Munshi2022}, but we do not expect a big difference in this case because the spatial frequencies of experimental imperfections (noise and contamination) do not necessarily hold specific phases in STEM images. Additionally, their spatial distributions are partially regularized by image discriminators as well.

\subsection*{CycleGAN training}
The training set of CycleGAN is composed by both experimental images and simulated STEM images. To prepare the CycleGAN training set, for each data set, we normalize all the image intensities into [-1, 1] within their group by their minimum and maximum value after a saturation step ($\pm\,3.5 \sigma$). We then cut them into $256 \times 256$ patches. The batch-size is 42, however when adding each patch into the batch, they are randomly augmented by a combination of rotation (multiples of 90 degrees) and flipping (vertically or horizontally). The out-of-the-box simulated images initially contain no noise, which is detrimental to a CycleGAN because a sufficient amount of variability is required to generate unique images. We add Gaussian noise (std = 0.1) to the out-of-the-box simulated images to ensure enough variability across the simulation data set. Note that this parameter does not require further manual optimization. All networks were trained for 297 epochs, where in the first 148 epochs, the learning rate was 0.032, and from epochs 148--297, the learning rate linearly decayed to zero. Supplemental Figure 5 shows the training loss as a function of epochs. One key difference between traditional CycleGAN training and here is that the FFT discriminator loss is also incorporated in the generator loss alongside the real-space discriminator loss.
For the discriminators such as $D_{X,\text{img}}$, and $D_{Y,\text{img}}$, the loss functions are 

\begin{align}
    \mathcal{L}(D_{X,\text{img}}) & = \mathbb{E}_{y\sim p_{\text{data}}(y)}||    D_{X,\text{img}}(F(y))||_{2}
                                    + \mathbb{E}_{x\sim p_{\text{data}}(x)}||1 - D_{X,\text{img}}  (x) ||_{2}, \\
    \mathcal{L}(D_{Y,\text{img}}) & = \mathbb{E}_{x\sim p_{\text{data}}(x)}||    D_{Y,\text{img}}(G(x))||_{2}
                                    + \mathbb{E}_{y\sim p_{\text{data}}(y)}||1 - D_{Y,\text{img}}  (y) ||_{2}
\end{align}

and similarly for $D_{X,\text{FFT}}$, and $D_{Y,\text{FFT}}$. For the generators $G: X \rightarrow Y$ and $F: Y\rightarrow X$, the loss functions are 

\begin{align}
\mathcal{L}(G) &= \mathcal{L}_{\text{adv}}(G) + \lambda_{\text{cyc}}\mathcal{L}_{\text{cyc}}(F,G) + \lambda_{\text{id}}\mathcal{L}_{\text{id}}(G)\\
\mathcal{L}(F) &= \mathcal{L}_{\text{adv}}(F) + \lambda_{\text{cyc}}\mathcal{L}_{\text{cyc}}(F,G) + \lambda_{\text{id}}\mathcal{L}_{\text{id}}(F)
\end{align}

Here, the adversarial loss $\mathcal{L}_{\text{adv}}$ that is commonly used in normal GANs~\cite{Goodfellow2014} is the term that aims to fool its respective discriminator(s):

\begin{equation}
\mathcal{L}_{\text{adv}}(G) = \mathbb{E}_{x\sim p_{\text{data}}(x)}||1 - D_{Y,\text{img}}(G(x))||_{2} 
                            + \mathbb{E}_{x\sim p_{\text{data}}(x)}||1 - D_{Y,\text{FFT}}(G(x))||_{2}
\end{equation}

The cycle-consistency loss is given by

\begin{equation}
\mathcal{L}_{\text{cyc}}(F,G) = \mathbb{E}_{x\sim p_{\text{data}}(x)}||F(G(x)) - x ||_{1} 
                              + \mathbb{E}_{y\sim p_{\text{data}}(y)}||G(F(y)) - y ||_{1}
\end{equation}

and the identity loss is given by 

\begin{align}
\mathcal{L}_{\text{id}}(G) = \mathbb{E}_{y\sim p_{\text{data}}(y)}||G(y) - y ||_{2}.
\end{align}

For the cycle-consistency loss, $\lambda_{\text{cyc}}$ was 10, and for the identity loss, $\lambda_{\text{id}}$ was 5. Each CycleGAN in Table~\ref{tab:FCNresults} has different numbers of 1K-resolution STEM images as their training set. CycleGAN-A has 107 images from Day A, CycleGAN-B has 211 images from Day B, and CycleGAN-AB has 318 images from Day AB. All the models are constructed with Tensorflow and ran on the Delta cluster at the National Center for Supercomputing Applications (NCSA). There, training a CycleGAN model from scratch takes around 6 hours to complete using a single NVIDIA A40 GPU.

\subsection*{FCN architecture and training}
We employ the same FCN architecture as in our previous report~\cite{Lee2020}. We used the Adam optimizer with a categorical cross entropy loss function. For each $1024 \times 1024$ (1K-resolution) STEM image that is used for FCN training, we cut the image into 16 $256 \times 256$ patches with stride (256, 256). Each of these patches is then rotated, flipped, and scale jittered, so that each $256\times256$ patch has 24 different augmentations. Hence, each 1K-resolution STEM image leads to $16 \times 24 = 384$ smaller patches, which are then divided into training set and simulated test set with a proportion of 10:1. In all FCN training, we use 107 1K-resolution STEM images. While training the FCN, each epoch randomly chooses 1,000 $256 \times 256$ patches from the training set. We train for 500 epochs.

\subsection*{FCN test sets}
The FCNs are tested with simulated (Supplemental Table 2) and experimental test sets (Table~\ref{tab:FCNresults}). The simulated test sets are generated along with the FCN training sets but are picked out before the FCN training. The experimental test sets are constructed by selecting 3 images from Day A (107 images) and Day B (211 images) experimental data set, respectively. The images are selected to roughly cover the entire experiment time span, which is usually 4--8 hours. The single Se vacancy positions are determined by the local contrast, as single vacancy in monolayer \ce{WSe2} would produce only half of the ADF-STEM pixel intensity comparing to a defect-free Se column. Each image is slightly Fourier filtered (high pass and masking the Bragg reflections) to aid the manually labeling process, while the final labeling is confirmed with the raw image. Test images are $21 \times 21$ nm\textsuperscript{2} large with $1024 \times 1024$ pixels, while test set A and B contains 117 and 177 single Se vacancies per image in average, respectively.


\section*{DATA AVAILABILITY}
The data sets are available on Zenodo: \url{https://doi.org/10.5281/zenodo.7696721}

\section*{CODE AVAILABILITY}
The codes are available on github: \url{https://github.com/ClarkResearchGroup/stem-learning}

\section*{ACKNOWLEDGEMENT}
This material is based upon work supported by the U.S. Department of Energy, Office of Science, Office of Basic Energy Sciences, Division of Materials Sciences and Engineering under award number DE-SC0020190, which supported the electron microscopy and related data analysis. We acknowledge Yue Zhang and Prof. Arend van der Zande for the \ce{WSe2} sample fabrication. This work was carried out in part in the Materials Research Laboratory Central Facilities at the University of Illinois Urbana--Champaign. This research is also part of the Delta research computing project, which is supported by the National Science Foundation (award OCI 2005572), and the State of Illinois. Delta is a joint effort of the University of Illinois Urbana--Champaign and its National Center for Supercomputing Applications.

\section*{AUTHOR CONTRIBUTIONS}
Under supervision by B.K.C., A.K. built and evaluated the CycleGAN and FCN models, and conducted the data set similarity analysis. Under supervision by P.Y.H., C.-H.L. acquired and analyzed the experimental and simulated STEM images, and conducted the data set similarity analysis. A.K. and C.-H.L. contributed equally to this work and are considered as co-first authors. All authors read and contributed to the manuscript.

\section*{COMPETING INTERESTS}
The authors declare no competing financial interest.

\section*{ADDITIONAL INFORMATION}
\noindent\textbf{Supplementary information} The online version contains supplementary material including extra figures of CycleGAN denoising, CycleGAN without FFT discriminators, cross-evaluated images, power spectra comparison, and training losses per epoch.\\
\noindent\textbf{Correspondence} and requests for materials should be addressed to Pinshane Y. Huang or Bryan K. Clark.

\printbibliography



\newpage
\section*{FIGURE LEGENDS}

\begin{table}
    \centering
    \begin{tabular}{|c|c|c|c|}
     \hline
     & graphene & \ce{WSe2} & \ce{SrTiO3} \\
     \hline
     Simulation & 32.69 & 31.87& 19.60 \\
     CycleGAN &  1.66 &  0.35&  0.47 \\
     \hline
    \end{tabular}
    \caption{FID scores of the simulated data set without noise and the CycleGAN-processed data set with respect to the experimental data set for graphene, \ce{WSe2}, and \ce{SrTiO3}. Note that individual CycleGANs are trained for each materials system.}
    \label{tab:FIDs}
\end{table}

\begin{table}
\centering
\begin{tabular}{|c||ccccccccc||}
\hline
\multirow{2}{*}{}                                                    & \mc{9}{c||}{\textbf{FCN test sets}}                                                                                                                                                                           \\ \cline{2-10} 
                                                                     & \mc{3}{c||}{A}                                                         & \mc{3}{c||}{B}                                                         & \mc{3}{c||}{AB}                                             \\ \hline
\tbf{\begin{tabular}[c]{@{}c@{}}FCN\\ training sets\end{tabular}}    & \mc{1}{c|}{P}        & \mc{1}{c|}{R}           & \mc{1}{c||}{F1}       & \mc{1}{c|}{P}        & \mc{1}{c|}{R}        & \mc{1}{c||}{F1}          & \mc{1}{c|}{P}           & \mc{1}{c|}{R}         & F1        \\ \hhline{|-||===||===||===||}
no noise                                                             & \mc{1}{c|}{66}       & \mc{1}{c|}{84}          & \mc{1}{c||}{74}       & \mc{1}{c|}{15}       & \mc{1}{c|}{83}       & \mc{1}{c||}{25}          & \mc{1}{c|}{22}          & \mc{1}{c|}{83}        & 35        \\ \hline
manual noise                                                         & \mc{1}{c|}{91}       & \mc{1}{c|}{89}          & \mc{1}{c||}{\tbf{90}} & \mc{1}{c|}{66}       & \mc{1}{c|}{\tbf{95}} & \mc{1}{c||}{78}          & \mc{1}{c|}{74}          & \mc{1}{c|}{\tbf{92}}  & \tbf{82}  \\ \hline
CycleGAN-A                                                           & \mc{1}{c|}{\cc87}    & \mc{1}{c|}{\cc\tbf{91}} & \mc{1}{c||}{\cc89}    & \mc{1}{c|}{12}       & \mc{1}{c|}{57}       & \mc{1}{c||}{20}          & \mc{1}{c|}{22}          & \mc{1}{c|}{71}        & 34        \\ \hline
CycleGAN-B                                                           & \mc{1}{c|}{97}       & \mc{1}{c|}{40}          & \mc{1}{c||}{56}       & \mc{1}{c|}{\cc85}    & \mc{1}{c|}{\cc79}    & \mc{1}{c||}{\cc\tbf{82}} & \mc{1}{c|}{87}          & \mc{1}{c|}{63}        & 73        \\ \hline
CycleGAN-AB                                                          & \mc{1}{c|}{\tbf{98}} & \mc{1}{c|}{49}          & \mc{1}{c||}{65}       & \mc{1}{c|}{\tbf{85}} & \mc{1}{c|}{73}       & \mc{1}{c||}{78}          & \mc{1}{c|}{\cc\tbf{89}} & \mc{1}{c|}{\cc63}     & \cc74     \\ \hline
\end{tabular}
\caption{FCN performance metrics with different training sets. 5 different FCN training sets are prepared, including simulation without noise, simulation with manually optimized noise, and CycleGANs trained on 3 different sets of experimental images (Day A, Day B, and Day AB). The FCN performance are tested with manually labeled experimental test sets (A, B, and AB) and evaluated by precision (P), recall (R), and F1 scores with units in percentage (\%). In the last three rows, the cells with matching CycleGAN training set and FCN test set are colored in gray; using a matching set between CycleGAN and FCN would be the standard approach for production runs. Unsurprisingly, running the CycleGAN and FCN on different sets of data (white boxes), which one would not do in practice, leads to poorer accuracy. The highest value of each column is bolded.}
\label{tab:FCNresults}
\end{table}

\end{document}


\newpage



\newpage

\begin{figure}[H]
    \includegraphics[width = 16cm]{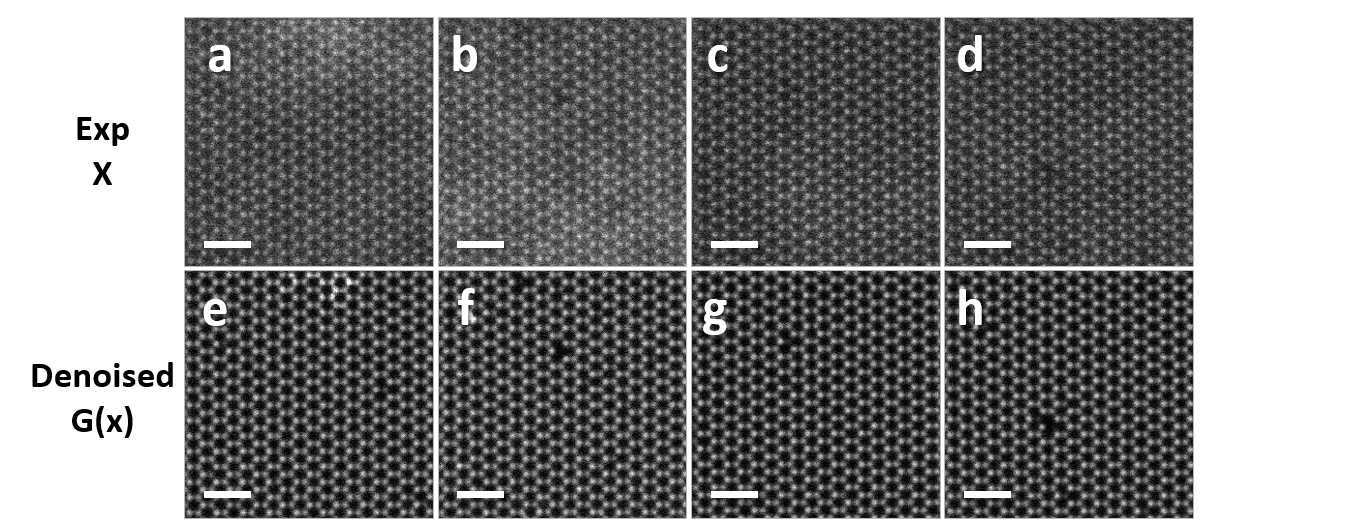}
    \centering
    \caption{Experimental ADF-STEM images of monolayer \ce{WSe2} (a--d) before and (e--h) after denoising by the CycleGAN generator $G$, where $G: X \rightarrow Y$. Note that the denoised images (e--h) still contain small amounts of noise, because Gaussian noise (std = 0.1) are manually added to the out-of-the-box simulated images before CycleGAN training to ensure enough variability in simulation domain $Y$ and the generator $G$ is therefore trying to replicate this Gaussian noise. Scale bars are equal to 1 nm.}
    \label{fig:S1}
\end{figure}

\begin{figure}[H]
    \includegraphics[width = 16cm]{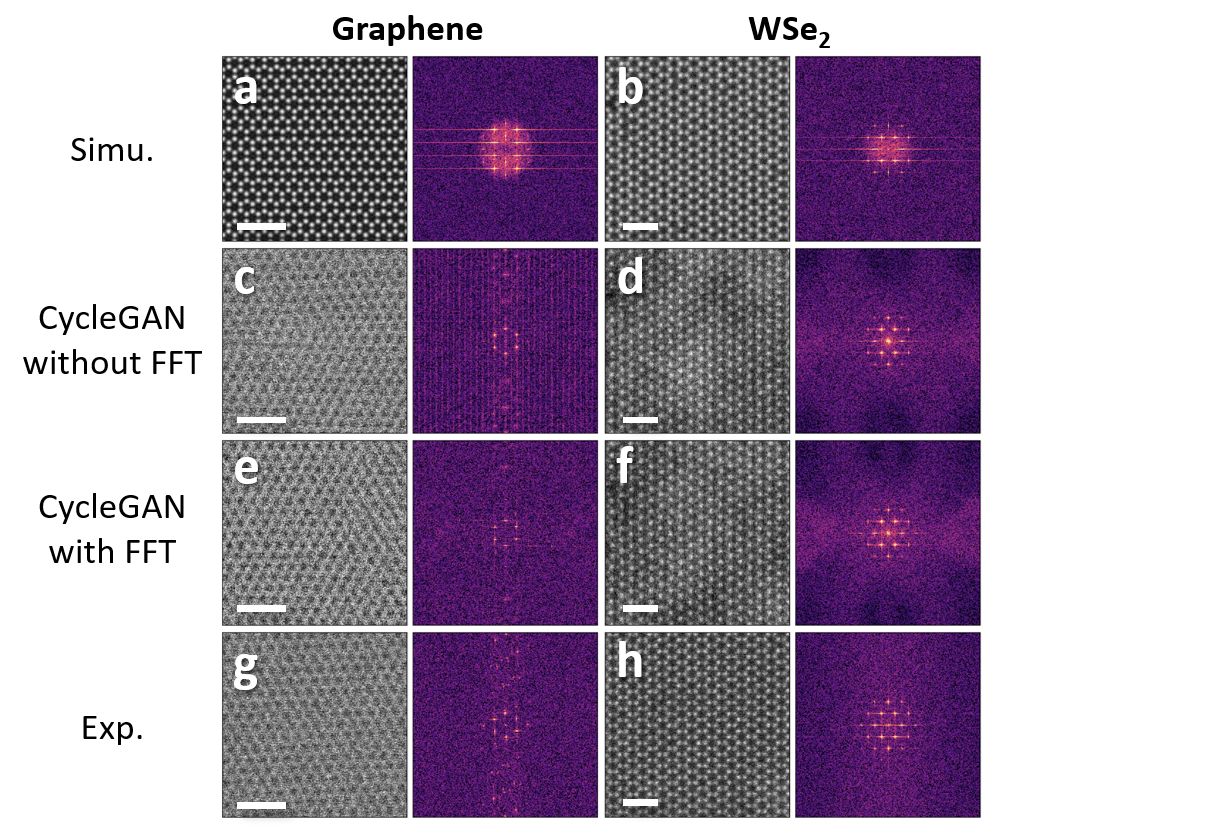}
    \centering
    \caption{Comparison of images of graphene and \ce{WSe2} from (a,b) simulation, (c,d) CycleGAN without FFT discriminator, (e,f) CycleGAN with FFT discriminator, and (g,h) experiment. Each row displays the real space images along with their power spectra. Power spectrum of the image is calculated by the fast Fourier transform (FFT) and displayed in log scale for clarity. While the addition of the FFT discriminators in the CycleGAN may not be easily noticeable in the case of \ce{WSe2}, it is critical while processing low signal-to-noise ratio images of graphene. For example, a CycleGAN without FFT discriminator generates graphene images with vertical, streaky artifacts clearly visible in the FFT and to a lesser extent in the real-space image shown in (c). The addition of FFT discriminator significantly reduces such artifacts in (e) and generates images with realistic k-space features. Scale bars are equal to 1 nm.}
    \label{fig:S2}
\end{figure}

\begin{figure}[H]
  \includegraphics[width = 16cm]{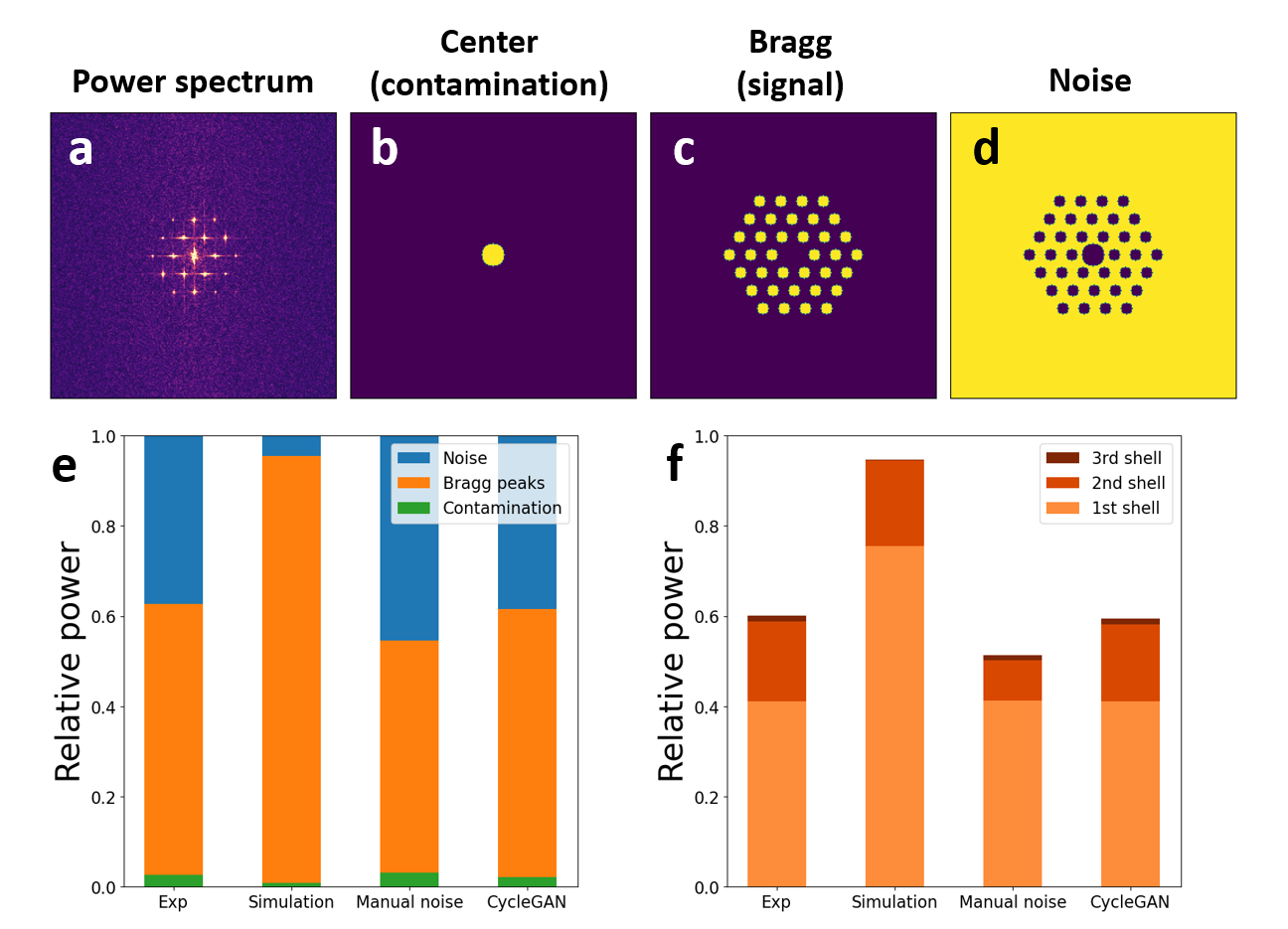}
  \caption{Quantitative comparison of data set similarities using their power spectra. (a) Representative power spectrum of the experimental data set of \ce{WSe2}. The intensity is displayed in log scale for clarity. Binary mask for (b) low frequency components (surface contamination), (c) Bragg peaks (signal from crystal structure), and (d) Non-Bragg peaks components (noise). The signal power of each component is integrated within the masked region. (e--f) Bar plots of relative power of each component, while (f) shows the power distribution of individual shells of the `Bragg peaks' in (e). These data show that the power distribution of the CycleGAN generated data is the best match for the experimental data.} 
  \label{fig:S3}
\end{figure}

\begin{figure}[H]
      \includegraphics[width = 12cm]{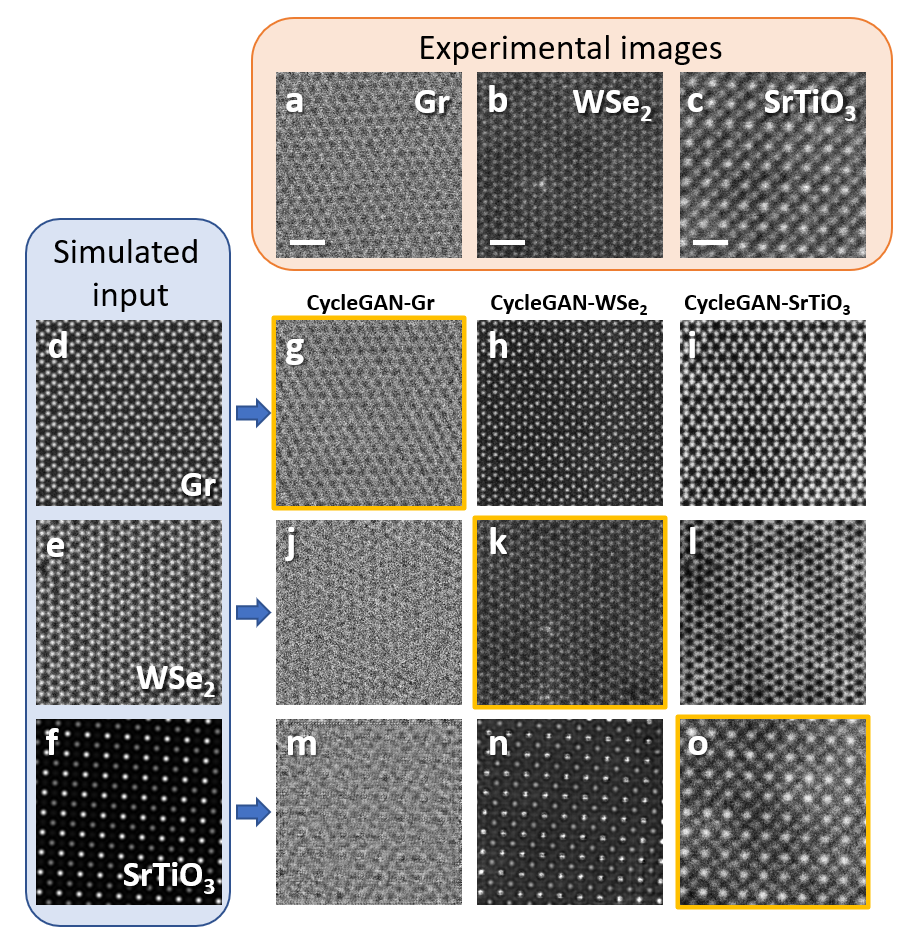}
      \centering
      \caption{Simulated STEM images processed by CycleGANs trained with different materials system. (a--c) Experimental STEM images of graphene, \ce{WSe2}, and \ce{SrTiO3}. Each image has different crystal structure and noise profile. These experimental images provide the targeted `style' for CycleGAN training. (d--f) Simulated STEM images that are used as the input for the following CycleGAN processing. (g--i) Simulated graphene image being individually processed by CycleGANs trained with different material systems. `CycleGAN-Gr' denotes the CycleGAN model is trained with experimental and simulated graphene images, so no large geometrical deformation is involved in the domain-to-domain transfer. (j--l) Simulated \ce{WSe2} image being processed by CycleGAN-Gr, CycleGAN-\ce{WSe2}, or CycleGAN-\ce{SrTiO3}, respectively. (m--o) Simulated \ce{SrTiO3} image being processed by each CycleGAN. These cross-evaluated images form a $3 \times 3$ matrix. The images with matching input and CycleGAN training data (g, k, and o) are marked with yellow outlines. These images are reproduced from Figure 1a--c in the main text. The off-diagonal elements (e.g. unmatch) of this matrix are images processed by CycleGAN trained with different materials system. These images perform poorly as expected, because the features transferred by the CycleGAN, which are the amount and type of noise as well as surface damage and contamination, would typically be different for different material systems. Besides, the translations were trained on dissimilar crystal structures that could have very different domain embeddings. Scale bars are equal to 1 nm.}
      \label{fig:S4}
    \end{figure}

\begin{figure}[H]
  \includegraphics[width=16cm]{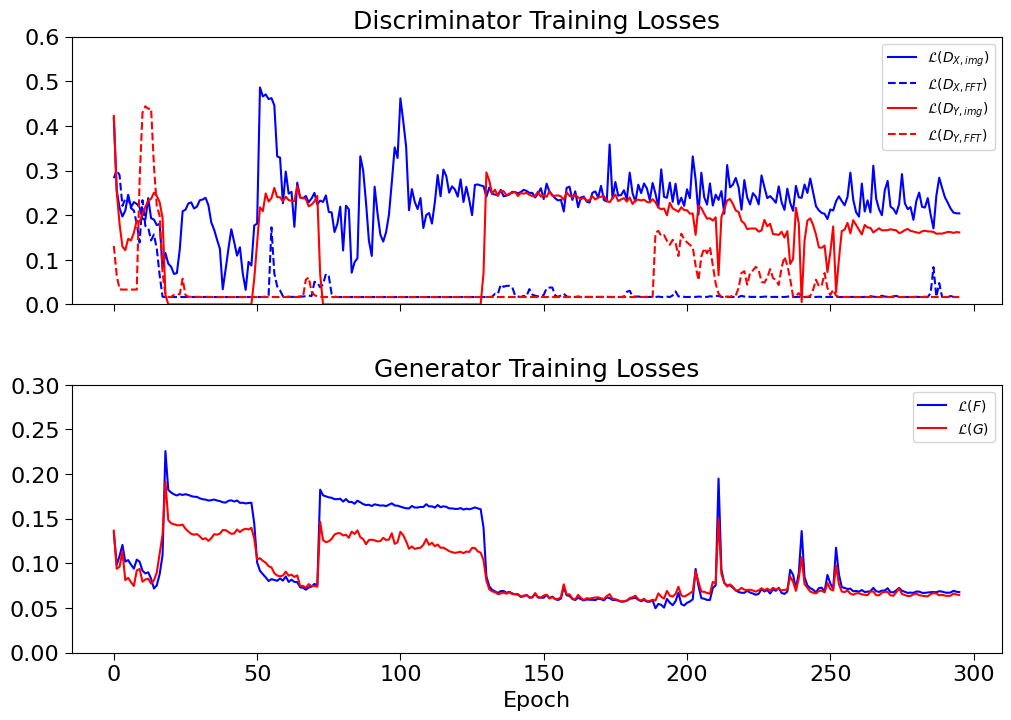}
  \caption{CycleGAN training loss versus number of epochs for experimental training set Day AB. The top plot shows the training losses for the two real-space discriminators $D_{Y, \text{img}}$, $D_{X,\text{img}}$ and the two reciprocal-space discriminators $D_{Y,\text{FFT}}$, $D_{X,\text{FFY}}$. The bottom plot shows training losses for the two generators $G : X\rightarrow Y$ and $F:Y\rightarrow X$. In the middle of training these losses fluctuate, but reach an equilibrium towards the end as expected since the discriminators and generators are competing against each other.}
  \label{fig:S5}
\end{figure}

\begin{table}[H]
   \centering
    \begin{tabular}{|c||ccc|}
    \hline
                               & \mc{3}{c|}{\textbf{FID reference image sets}}    \\ \hline
    \textbf{FCN training sets} & \mc{1}{c|}{A}      & \mc{1}{c|}{B}      & AB     \\ \hline
    no noise                   & \mc{1}{c|}{31.864} & \mc{1}{c|}{32.256} & 31.992 \\ \hline
    manual noise               & \mc{1}{c|}{0.727}  & \mc{1}{c|}{1.175}  & 0.930  \\ \hline
    CycleGAN-A                 & \mc{1}{c|}{\cc0.411}  & \mc{1}{c|}{1.414}  & 0.990  \\ \hline
    CycleGAN-B                 & \mc{1}{c|}{0.808}  & \mc{1}{c|}{\cc1.110}  & 0.914  \\ \hline
    CycleGAN-AB                & \mc{1}{c|}{\tbf{0.361}}  & \mc{1}{c|}{\tbf{0.722}}  & \cc\tbf{0.499}  \\ \hline
    \end{tabular}
   \caption{FID scores of the 5 different FCN training sets with the 3 FCN test sets A, B, and AB. Smaller FID scores represent a higher similarity with respect to the test sets, hence suggesting better training data. The cells with matching CycleGAN training set and FCN test set are colored in gray. The lowest FID score of each column is bolded.}
   \label{tab:S1}
\end{table}

\begin{table}[H]
    \centering
    \begin{tabular}{|c||c|c|c|}
    \hline
    \textbf{FCN training sets} & P & R & F1 \\ \hline \hline
    no noise                   & 99.8  & 99.6  & 99.7   \\ \hline
    manual noise               & 99.2  & 96.0  & 97.6   \\ \hline
    CycleGAN-A                 & 99.5  & 98.2  & 98.8   \\ \hline
    CycleGAN-B                 & 99.5  & 97.8  & 98.6   \\ \hline
    CycleGAN-AB                & 99.9  & 98.4  & 99.1   \\ \hline
    \end{tabular}
    \caption{FCN performance metrics evaluated with simulated test sets generated along with each training set. These values are higher than the FCN performance on experimental datasets and are provided for reference because evaluating on simulated data is common in the literature. As shown in this table, the FCN performance on simulated data is quite high, above 96 for each metric. }
    \label{tab:S2}
\end{table}


\begin{table}[H]
    \centering
    \begin{tabular}{|c||cccccc||ccc|}
    \hline
    \multirow{2}{*}{}                                                    & \mc{9}{c|}{\textbf{FCN test sets}}                                                                                                                                                                                 \\ \cline{2-10} 
                                                                         & \mc{3}{c||}{A}                                                                                 & \mc{3}{c||}{B}                                                                                 & \mc{3}{c|}{AB}     \\ \hline
    \textbf{\begin{tabular}[c]{@{}c@{}}FCN\\ training sets\end{tabular}} & \mc{1}{c|}{P}           & \mc{1}{c|}{R}           & \mc{1}{c||}{F1}          & \mc{1}{c|}{P}           & \mc{1}{c|}{R}           & \mc{1}{c||}{F1}          & \mc{1}{c|}{P}        & \mc{1}{c|}{R}        & F1       \\ \hline
    CycleGAN-A1                                                          & \mc{1}{c|}{\cc89}       & \mc{1}{c|}{\cc90}       & \mc{1}{c||}{\cc\tbf{89}} & \mc{1}{c|}{22}          & \mc{1}{c|}{84}          & \mc{1}{c||}{35}          & \mc{1}{c|}{32}       & \mc{1}{c|}{87}       & 46       \\ \hline
    CycleGAN-A2                                                          & \mc{1}{c|}{\cc77}       & \mc{1}{c|}{\cc\tbf{93}} & \mc{1}{c||}{\cc84}          & \mc{1}{c|}{10}       & \mc{1}{c|}{71}          & \mc{1}{c||}{18}          & \mc{1}{c|}{18}       & \mc{1}{c|}{79}       & 29       \\ \hline
    CycleGAN-A3                                                          & \mc{1}{c|}{\cc84}       & \mc{1}{c|}{\cc92}       & \mc{1}{c||}{\cc88}          & \mc{1}{c|}{11}       & \mc{1}{c|}{46}          & \mc{1}{c||}{18}          & \mc{1}{c|}{23}       & \mc{1}{c|}{64}       & 33       \\ \hline
    CycleGAN-B1                                                          & \mc{1}{c|}{\tbf{99}}    & \mc{1}{c|}{40}          & \mc{1}{c||}{57}          & \mc{1}{c|}{\cc82}       & \mc{1}{c|}{\cc81}       & \mc{1}{c||}{\cc81}       & \mc{1}{c|}{86}       & \mc{1}{c|}{65}       & \tbf{74} \\ \hline
    CycleGAN-B2                                                          & \mc{1}{c|}{98}          & \mc{1}{c|}{28}          & \mc{1}{c||}{44}          & \mc{1}{c|}{\cc\tbf{85}} & \mc{1}{c|}{\cc79}       & \mc{1}{c||}{\cc\tbf{82}} & \mc{1}{c|}{\tbf{87}} & \mc{1}{c|}{59}       & 70       \\ \hline
    CycleGAN-B3                                                          & \mc{1}{c|}{92}          & \mc{1}{c|}{81}          & \mc{1}{c||}{86}          & \mc{1}{c|}{\cc28}       & \mc{1}{c|}{\cc\tbf{85}} & \mc{1}{c||}{\cc42}       & \mc{1}{c|}{38}       & \mc{1}{c|}{\tbf{83}} & 52       \\ \hline
    \end{tabular}
    \caption{FCN performance metrics with training sets processed by different CycleGANs, where each CycleGAN is trained with only 6 experimental 1K-resolution images. CycleGAN-A1 to A3 are the images generated by CycleGANs trained with subsets that are selected from the original Day A data set, while B1 to B3 are generated by CycleGANs trained with subsets from the Day B. Each subset has 6 images that are chosen by the proximity of image acquisition time with respect to the manually labeled test images in each data set. The subset images are selected by acquisition time to emulate the effect of time-dependent microscope instabilities. Images that are acquired around the same time are considered to have similar microscope conditions. The FCN performance are tested with manually labeled experimental test sets (A, B, and AB) and evaluated by precision (P), recall (R), and F1 scores with units in percentage (\%). The cells with matching CycleGAN training set and FCN test set are colored in gray; using a matching set between CycleGAN and FCN would be the standard approach for production runs. Unsurprisingly, if you run the CycleGAN and FCN on different sets of data (the non-grayed boxes), which one would not do in practice, the accuracy is much worse. The cells with matching CycleGAN training set and FCN test set are colored in gray. The highest value of each column is bolded.}
    \label{tab:S3}
\end{table}
